\documentclass[12pt]{article}
\usepackage{epsfig}

\parskip=\medskipamount 
plus.3em minus.5em \abovedisplayshortskip=.5em plus.2em minus.4em
\renewcommand{\thesection}{\arabic{section}}

\catcode`@=11

\@addtoreset{equation}{section} \@addtoreset{equation}{subsection}
\def\theequation{\ifnum\value{section}=0 \arabic{equation}\ignorespaces
\else \ifnum\value{section}=-1 A.\arabic{equation}\ignorespaces
\else \ifnum\value{subsection}=0
\thesection.\arabic{equation}\ignorespaces \else
\thesection.\arabic{subsection}.\arabic{equation}\ignorespaces
                             \fi
                        \fi
                   \fi}

{\catcode`\'=\active \def'{{}^\bgroup\prim@s}}

\catcode`@=12

\newcommand{\bq}{\begin{equation}}
\newcommand{\be}{\begin{equation}}
\newcommand{\fq}{\end{equation}}
\newcommand{\ee}{\end{equation}}
\newcommand{\bqr}{\begin{eqnarray}}
\newcommand{\beqs}{\begin{eqnarray}}
\newcommand{\fqr}{\end{eqnarray}}
\newcommand{\eeqs}{\end{eqnarray}}

\newcommand{\rf}[1]{(\ref{#1})}

\def\bop#1{\setbox0=\hbox{$#1M$}\mkern1.5mu
    \vbox{\hrule height0pt depth.04\ht0
    \hbox{\vrule width.04\ht0 height.9\ht0 \kern.9\ht0
    \vrule width.04\ht0}\hrule height.04\ht0}\mkern1.5mu}

\begin{document}
\thispagestyle{empty}

\begin{flushright}
\begin{tabular}{l}
hep-th/0509081 \\
\end{tabular}
\end{flushright}

\vskip .6in
\begin{center}

{\bf Novel FIR Inversion with Only FIRs}

\vskip .6in

{\bf Gordon Chalmers}
\\[5mm]

{e-mail: gordon@quartz.shango.com}

\vskip .5in minus .2in

{\bf Abstract} 

\end{center}

\noindent
The inversion of an FIR data sampling is usually stated to be possible with 
the use of a potentially unstable IIR, and in particular circumstances.  It 
is possible to accomplish the same inversion with the doubling of an FIR 
sampling and with only FIRs for the sampling and the inversion.  This note 
presents the configuration, which appently is not in the literature, for 
perfect signal reconstruction.

\vfill\break

\vskip .2in

The sampling of information with FIR filters is very common 
\cite{filterexample}.  The signal 
reconstruction of a data stream sampled with these filters is typically not 
analyzed in a simple manner, and the invocation of a potentially unstable IIR 
filter bank is utilized.  It is commonly stated that the inversion of a signal 
with only FIRs is not possible.  This is not the case, and in this note a 
signal reconstruction is presented that performs this task without loss of 
information.  

A digital data stream is denoted by $X(n)$, and a k-tap filtering is 
performed with the transform, 

\bqr 
Y(n)=\sum_{i=1}^k b_i X(n-i) \ .  
\fqr 
The coefficients $b_k$ are real for a real data stream, but the duplication 
of the process can be performed with complex coefficients $c_k$.  In the 
latter scenario, the real data stream $X(n)$ is used and the real part of 
$Y(n)$ is taken, i.e. the FIR filtering is identical to taking ${\cal R}e ~
Y(n)$ with the real parts of the complex coefficients ${\cal R}e ~c_k=b_k$. 

The duplication of the FIR filtering is not required in general.  However, 
with this process of using complex coefficients $c_k$, an perfect 
reconstruction of the filtered signal $Y(n)$ can be achieved with using an 
additional complex FIR, with suitably chosen taps.  The use of an IIR is not 
required.  

The double process of the two FIR filterings results in the signal, 

\bqr 
Z(n)=\sum_{i=1}^k \sum_{j=1}^k d_k c_k X(n-i-j) \ .  
\fqr 
Choosing the complex coefficients $d_k$ appropriately results in $Z(n)=X(n)$.  

Consider the example of a 4-tap filter.  The reconstruction appears with the 
following formulae, 

\bqr 
c_0 d_0 = \rho + i \alpha_1  
\cr 
c_1 d_0 + c_0 d_1 = i\alpha_2 
\cr 
c_1 d_1 + c_0 d_2 + c_2 d_0 = i \alpha_3 
\cr 
c_2 d_1 + c_1 d_2 + c_3 d_0 + c_0 d_3 = i \alpha_4
\cr
c_2 d_2 + c_3 d_1 + c_1 d_3 = i\alpha_5 
\cr 
c_3 d_2 + c_2 d_3 = i \alpha_6 
\cr 
c_3 d_3 = i \alpha_7  \ . 
\label{inverse}
\fqr 
In general these equations are not invertible for real inputs $c_i$ and $d_i$, 
and with $\alpha_j=0$.  Taking these taps to be complex, and with $\alpha_i$ 
arbitrary, allows for a general solution.  Then the signal 
${\cal R}e ~Z(n)=X(n)$.  

The case of a DFT is known, pertaining to $\vert c_i\vert=1$, that is, with 
coefficients on the unit circle.  More general filtering requires these 
coefficients to be anywhere in the complex plane.  

The requirements to invert a real signal are then ${\cal R}e ~c_i=b_i$ and the 
solution to the general system in \rf{inverse}.  The coefficients $\alpha_i$ 
can be anything, and $\rho$ is a parameter than may rescale the output.  The 
system in \rf{inverse} has seven complex equations, in which only six real 
components are non-trivial as the $\alpha_i$ are a priori free parameters 
(they can be chosen to solve the system of equations).  There are only four 
inputs, the $b_i$.  There are twelve real unknowns ${\cal I}m ~c_i$ and 
$d_i$.  

The conditions in \rf{inverse} form a matrix equation, 

\bqr 
\pmatrix{ c_0  &  0    &  0    &  0    \cr 
          c_1  &  c_0  &  0    &  0    \cr 
          c_2  &  c_1  &  c_0  &  0    \cr 
          c_3  &  c_2  &  c_1  &  c_0  \cr  
          0    &  c_3  &  c_2  &  c_1  \cr 
          0    &  0    &  c_3  &  c_2  \cr 
          0    &  0    &  0    &  c_3  \cr}   
\pmatrix{ d_0 \cr d_1 \cr d_2  \cr d_3}  =  i 
\pmatrix{ \alpha_1 \cr \alpha_2 \cr \alpha_3 \cr 
          \alpha_4 \cr \alpha_5 \cr \alpha_6 \cr \alpha_7 }  \ . 
\label{matrixequation}
\fqr 
The inversion of the algebraic system generates the complex $d_i$ 
parameters and the imaginary components of $c_i$ (the $b_i$).  For example, 
Matlab can used to invert the non-square matrix, followed by the 
multiplication with a vector of $\alpha_i$ entries and solve for the $d_i$.  
In general the $\alpha_i$ are non-vanishing; a set of zero entries gives 
zero values for the $d_i$ taps.   The 7 equations modeling the 
real components of \rf{matrixequation} can be written in a 8x8 form with the 
real and imaginary components of $d=d_r + id_i$.  This system is complete 
(7 equations in eight variables) and can be used to solve for the four 
$d_i$ in terms of the complex components of $c_i$.  The solutions to $d_i$ 
are then used to determine the $\alpha_i$ parameters.  

This procedure is straigtforward to implement for a general tap filter, 
even for large numbers.  The perfect inversion of the FIR filtered signal 
is accomplished without the use of a potentially unstable IIR on some data, 
and with perfect reconstruction.  The simple configuration requires the 
doubling of the input signal into real and imaginary 
parts, with another pair of filters to reconstruct.  The configuration 
is also more cost effective than using IIRs.

\end{document}